\documentclass[11pt,dvips]{article}
\usepackage{graphicx}
\usepackage{amssymb}
\usepackage{latexsym}

\begin{document}
\begin{titlepage}
\setcounter{page}{1}
\renewcommand{\thefootnote}{\fnsymbol{footnote}}

\begin{flushright}
ucd-tpg:09-06~~~~~\\
%arXiv:0911.xxxx
\end{flushright}

\vspace{5mm}
\begin{center}

 {\Large \bf
Symplectic Fluctuations for Electromagnetic Excitations\\ of Hall Droplets}
%ELECTROMAGNETIC EXCITATIONS OF LANDAU SYSTEMS
%VIA SVEDBERG WITTED MAP AND APPLICATION TO HALL DROPLETS}

\vspace{0.5cm}

{\bf Mohammed Daoud}$^a${\footnote {\it Facult\'e des Sciences,
D\'epartement de Physique, Agadir, Morocco; email: {\sf
daoud@pks.mpg.de}}}, {\bf Ahmed Jellal}$^{b,c,d}$\footnote{{\sf
jellal@pks.mpg.de} - {\sf jellal@ucd.ma}}  and  {\bf Abdellah Oueld
Guejdi}$^e$

\vspace{0.5cm}

$^a${\em  Max Planck Institute for Physics of Complex Systems,
N\"othnitzer Str. 38,\\
 D-01187 Dresden, Germany}
%\vspace{0.2cm}

{$^b$\em Physics Department, College of Sciences, King Faisal University,\\
PO Box 380, Alahsaa 31982,
Saudi Arabia}

{$^c$\em Saudi Center for Theoretical Physics, Dhahran, Saudi Arabia}

$^d${\em  Theoretical Physics Group, Faculty of Sciences,
 Choua\"ib Doukkali University,\\
PO Box 20, 24000 El Jadida, Morocco} %\vspace{0.2cm}

$^e${\em Department of Mathematics, Faculty of Sciences, University Ibn Zohr,\\
PO Box 8106, Agadir,
Morocco}%\\[1em]

\vspace{3cm}

\begin{abstract}

We show that the integer quantum Hall effect systems in plane, sphere or disc,
can be formulated in terms of an algebraic unified scheme.
This can be achieved by making use of
a generalized Weyl--Heisenberg algebra and investigating its basic features.
We study the electromagnetic excitation
and derive the Hamiltonian for droplets of fermions
on a two-dimensional Bargmann space (phase space). This
excitation is introduced through a deformation (perturbation) of the symplectic
structure of the phase space. We show  the major role of Moser's lemma in
dressing procedure, which allows us to eliminate the fluctuations of the
symplectic structure. We discuss the emergence of the Seiberg--Witten map and
 generation of an abelian noncommutative gauge field in the theory. As illustration
of our model, we give the action describing the electromagnetic excitation of a quantum Hall droplet
in  two-dimensional manifold.

\end{abstract}
\end{center}
\end{titlepage}

\newpage

%%%%%%%%%%%%%%%%%%%%%%%%%%%%%%%%
\section{Introduction}
%%%%%%%%%%%%%%%%%%%%%%%%%%%%%%%%%%%

The notion of deformation, which is very familiar in mathematics and
physics, is based on the philosophy of deforming suitable mathematical structures
behind the physical theories (e.g., complex, symplectic or algebraic structures). In this connection,
the quantum and relativistic mechanics are commonly regarded as $\hbar$-deformation and $1/c$-deformation,
respectively, of classical mechanics. Quantum algebras ($q$-deformation),
string theory ($\alpha'$-deformation) and noncommutative field theory ($\theta$-deformation)
are the most studied deformed theories during the last decades. The deformation leads often to a radical
changes of the parent theory inducing new physics as for instance the wave-particle duality in quantum mechanics
and $T$-duality in string theory. In this spirit and
motivated by string theory arguments [1], the idea of
noncommutativity at small length scales [2] has been drawn much
attention in various fields, see for instance [3,4].
The noncommutative field theory continues  to be investigated extensively.

In this paper,
we shall first define a generalized Weyl--Heisenberg algebra $W_{\kappa}$, which can be seen as a deviation
from the ordinary boson algebra. This generalization can
be understand in the context of the quantum algebras. %($q$-deformation)
It  can also be related to the concept of deformed boson algebra  introduced in the seventies [5], see also [6].
The $\kappa$-dependent algebra generalizes the usual boson algebra,
which corresponds to the case $\kappa = 0$. This generalized
algebra offers an advantage to carry out a simultaneous study of planar ($\kappa = 0$),
spherical ($\kappa = -1$) and hyperbolic ($\kappa = 1$) systems.
Using a quantum--classical correspondence, we construct the phase (Bargmann)  space equipped with a symplectic
structure encoding the dynamics of the system whose relevant symmetry is described by
the algebra $W_{\kappa}$.

The second facet of this work deals with the $\theta$-deformation. This is done by modifying
the symplectic structure given by a two-form $\omega_0$ of the Bargmann space.
This provides us with a nice tool to study the
electromagnetic excitations of
fermions living in plane, sphere or disc. This procedure is well-known in symplectic
geometry and it is deeply related to the noncommutative geometry. It found interesting applications
in quantum mechanics and planar quantum Hall effect, see for instance [7-11].
In fact, the interaction of a charged particle can be described
in a Hamiltonian formalism without a choice of a potential. The interaction can be
introduced through a two-form $F$ inducing fluctuations of the
parent symplectic form.

The modified symplectic two-form can be always rewritten as $\omega_0$
in a new coordinates system in the phase space. This can be done by using the celebrated
 Darboux transformation  for a flat phase space geometry and
for some particular form of the electromagnetic field $F$.
%In this work,
We give the explicit form of the dressing transformation, which eliminates the fluctuations of the symplectic
form $\omega_0$ and transforms $\omega_0 + F \to \omega_0$ in curved phase space.
This is a refined version of Darboux transformation recognized
in the mathematical literature as Moser lemma [12]. Interestingly, this transformation turns out
identical to Seiberg--Witten map [1]. As by product, we show that
the dynamics of the system becomes described by a Hamiltonian involving
terms encoding the effect of perturbation $F$ and a Chern--Simons like interaction.

On the other hand, it is well-known that the planar fermions in a strong
magnetic field are confined in the lowest Landau levels and behave
like a rigid droplet of liquid. This is the incompressible quantum
fluid picture proposed by Laughlin [13]. This led to new perspectives for
using the noncommutative geometry ideas to discuss the
quantum Hall phenomenon in the plane [14,15] and  developing
later generalizations to other geometries with arbitrary
dimensions [16-26]. Therefore, we give an illustration of our analysis by
studying
the effect of the symplectic fluctuations on the quantum Hall droplets and comparing
the obtained results with ones  recently presented in the literature [27].

The outline of the paper is as follows. In section 2,
we introduce a generalized Weyl--Heisenberg algebra. We discuss
the corresponding Hilbertian representation and the analytical Bargamann
realization. It is remarkable that the obtained realization leads
to the Klauder--Perelomov coherent states. Also, it is important
to stress that this generalization induces, at geometrical level,
the curvature of the Bargmann space. The flat geometry
is recovered in the limit of the standard Weyl--Heisenberg algebra.  Using the algebraic structure
of the generalized Weyl algebra, we
discuss the quantum mechanics on coset spaces ${\bf S}_{\kappa}^2$ in section 3.
This provides us with an unified scheme to quantum mechanically deal with
 spherical and
hyperbolic systems. Using the standard tools of geometric quantization,
we establish one to one correspondence between the lowest energy levels (the vacuum) wavefunctions
and the holomorphic sections (coherent states) defining the Bargmann space.
In section 4, we equip the Bargmann space (phase space) with a star product and
consequently we define the Moyal brackets. We also
give the potential lifting the degeneracies of the vacuum. This is helpful in the semi-classical
analysis done in the next sections.
In section 5, we discuss the noncommutative dynamics in Bargmann space, which
is achieved by modifying the associated symplectic structure. We show that the
Moser's lemma offers a nice way to eliminate the fluctuation (modification) of the symplectic
two-form. This can be done by making use of a dressing transformation similar
to the Darboux one for flat phase spaces.
The effect of the perturbation is then incorporated in the Hamiltonian.
We also show that the Moser dressing transformation induces in a non trivial way a
noncommutative gauge field and it coincides with the Seiberg--Witten map. As illustration,
we derive, in section 6, the semiclassical
 effective action describing  the electromagnetic excitation of
  two-dimensional quantum Hall droplet.
Finally, we conclude with
a summary of our results.

%%%%%%%%%%%%%%%%%%%%%%%%%%%%%%%%%%%%%%%%%%%%%%%%%%%%%%%%%%%%
\section{Generalized Weyl--Heisenberg algebra $W_{\kappa}$}
%%%%%%%%%%%%%%%%%%%%%%%%%%%%%%%%%%%%%%%%%%%%%%%%%%%%%%%%%%%%
The first step in our program is to introduce a generalized
Weyl--Heisenberg algebra. We give the corresponding Hilbertian representation and
we provide the analytical Bargmann realization.

%%%%%%%%%%%%%%%%%%%%%%%%%%%%%%%%%%%%%%%%%%%%%%%%%%%%%
\subsection{Algebraic structures}
%%%%%%%%%%%%%%%%%%%%%%%%%%%%%%%%%%%%%%%%%%%%%%%%%%%%

Let us consider an algebra $W_{\kappa}$ characterized by four
generators $x_+$, $x_-$ , $x_0$ and ${\mathbb I}$. They satisfy the
commutation relations
\begin{equation}\label{1}
[x_- , x_+] = \mathbb{I} + 2\kappa x_0 , {\hskip 1cm } [x_0 , x_+] = x_+, {\hskip 1cm }[x_0 , x_-] = -x_-{\hskip 1cm }
[ ~.~ , ~\mathbb{I} ] = 0
\end{equation}
where $\kappa$ a real parameter. It is clear that, the operator ${\mathbb I}$ belongs to the center  of $W_{\kappa}$.
When $\kappa$  is non null, then it can be rescaled to $+1$ or $-1$, hence we have three representatives
values of $\kappa = -1, 0, +1$.
For $\kappa = 0$, we have the usual
harmonic oscillator algebra.
The sign of this parameter
plays an important role in specifying the representation dimensions
associated with $W_{\kappa}$ as it will be discussed through this section.
Note that, this algebra is relevant
in the theory of exactly solvable potentials in one-dimension [28] and
 fractional supersymmetric quantum mechanics [29]. Indeed, in the context of
 a quantum system evolving in one-dimensional space,  $x_+$, $x_-$ and $x_0$ can
physically be interpreted as generalized creation, annihilation and number operators, respectively.
 It is
also important to stress that,  $W_{\kappa}$ can be realized
as a class of nonlinear oscillator
algebras through the so-called deformed structure function [6]. Accordingly, the generators  $x_+$, $x_-$ and $x_0$ can be realized as
\begin{equation}\label{2}
x_+ = a_+~ \sqrt{{\mathbb I} + \kappa a_+a_-}, \qquad
x_- = \sqrt{{\mathbb I} + \kappa a_+a_-} ~a_-  , \qquad  x_0 = a_+a_-
\end{equation}
in terms of the ordinary creation and annihilation operators of the harmonic oscillator algebra.
Finally, we mention that $W_{\kappa}$ provides an unified scheme to deal with the
planar, spherical and hyperbolic geometries as it will be explained in the next section.

%%%%%%%%%%%%%%%%%%%%%%%%%%%%%%%%%%%%%%%%%%
\subsection{Hilbertian representation }
%%%%%%%%%%%%%%%%%%%%%%%%%%%%%%%%%%%%%%%%%

A Hilbertian representation corresponding to
the algebra $W_{\kappa}$ can be defined as follows. Let us denote by
${\cal F}= \{ \vert s , n \rangle , n = 0, 1, 2, \cdots , d({\kappa})\}$
the Hilbert--Fock space where the generators
 $x_+$, $x_-$,  $x_0$ and ${\mathbb I}$ act on.
 % as
 %$${\cal F} = \{ \vert s , n \rangle , n = 0, 1, 2, \cdots , d({\kappa})\}.$$
 The actions of the elements  ${\mathbb I}$ and $x_0$ are defined by
 \begin{equation}\label{3}
{\mathbb I} \vert s, n \rangle = 2s \vert s,  n \rangle, \qquad x_0 \vert s, n \rangle = n \vert s,  n \rangle
 \end{equation}
 and for the remaining generators, we have
\begin{equation}\label{4}
x_+ \vert s,  n \rangle = \sqrt{f_s(n+1)}\vert s, n+1 \rangle,\qquad
  x_- \vert s,  n \rangle = \sqrt{f_s(n)}\vert s, n-1 \rangle
 \end{equation}
 where the condition $x^- \vert 0 \rangle = 0 $ is considered.  The
 parameter $s$ is characterizing the  $W_{\kappa}$ irreducible representations and
 for the simplicity, we assume $2s \in {\mathbb N}^{\star}$ in the forthcoming analysis.
 Using the commutation relations (\ref{1}) and   actions (\ref{3}-\ref{4}), one
 can check that the functions $f_s(n)$ verify the recurrence relation
 \begin{equation}\label{5}
 f_s(n+1) - f_s(n) = 2s + 2\kappa n
  \end{equation}
 implemented by the condition $f_s(0) = 0$. A simple iteration of (\ref{5}) gives
 \begin{equation}\label{6}
 f_s(n) = 2sn + \kappa n(n-1).
 \end{equation}
 This structure function must be positive and therefore leads to the condition
 %\begin{equation}\label{7}
 $2s + \kappa (n-1) > 0$
 %\end{equation}
 for any quantum number $n>0$. This determines the dimension $d({\kappa})$ of the irreducible representation space ${\cal F}$.
 Indeed, for $\kappa = + 1$,  ${\cal F}$ is infinite
  dimensional, i.e. $d({\kappa}>0)= +\infty$, and for
 $\kappa = -1 $, the dimension of ${\cal F}$ is finite, i.e.
 $ n = 0, 1, 2, \cdots, 2s.$
Recall that, for $\kappa = 0$, we have the usual infinite dimensional bosonic Fock space.
Beside the Fock representations of the algebra $W_{\kappa}$, one can construct an analytical realization
 of the representation space ${\cal F}$. %This will be the subject of the next subsection.

%%%%%%%%%%%%%%%%%%%%%%%%%%%%%%%%%%%%%
\subsection{Bargmann realization}
%%%%%%%%%%%%%%%%%%%%%%%%%%%%%%%%%%%%%

The Bargmann realization associated with the algebra $W_{\kappa}$
 uses a suitably
defined Hilbert space of entire analytical functions.
We represent the Hilbert--Fock
 states $\vert s , n \rangle$
as power of complex variable $z$, such that
\begin{equation}\label{8}
| s, n \rangle  \longrightarrow C_{s, n} z^n.
\end{equation}
The  generator $x_-$ can be realized as first order
differential operator with respect to $z$. This is
\begin{equation}\label{9}
x_- \longrightarrow \frac{\partial}{\partial z}.
\end{equation}
Using the action of the annihilation operators on %the Fock space
${\cal F}$ and the correspondences (\ref{8}-\ref{9}), we show that
the coefficients $C_{ s, n}\ ( n > 0)$ take the form
\begin{equation}\label{10}
C_{s, n}= \frac{1} {\sqrt{n!}}\sqrt{(2s+\kappa(n-1))(2s+\kappa(n-2))\cdots (2s+\kappa)2s}\ C_{s, 0}
\end{equation}
and for convenience we set $C_{s, 0} = 1$. It is easy to see that the differential
realization of operator $x_0$ is given by
\begin{equation}\label{11}
x_0 \longrightarrow z\frac{\partial}{\partial z}.
\end{equation}
To achieve the present realization, we give the differential action of the
operator $x_+$. Indeed, by using its actions on the Hilbert--Fock space together with the recursion relation (\ref{10}),
we find
\begin{equation}\label{12}
x_+ \longrightarrow 2s z + \kappa z^2\frac{\partial}{\partial z}.
\end{equation}
Clearly,
the $W_{\kappa}$ generators act as first order linear differential
operators.

An  arbitrary vector $|\psi \rangle =
\sum_{n}\psi_{n}| s,  n\rangle$ of  ${\cal F}$
 can be mapped as
\begin{equation}\label{13}
\psi(z) =
\sum_{n} \psi_{n} C_{s,n} z^{n} .
\end{equation}
The inner product of two functions $\psi$ and $\psi'$ is defined by
\begin{equation}\label{14}
\langle\psi'|\psi\rangle = \int d^2z \Sigma(s;\bar z\cdot z)\ \psi'^{\star}(z)\  \psi(z).
\end{equation}
The integration measure $\Sigma$, assumed
to be isotropic, can explicitly be determined by choosing two functions $|\psi\rangle = |
s, n\rangle $ and $|\psi'\rangle = | s, n' \rangle $. A direct calculation
 shows that it can be cast, in compact
form, as
\begin{equation}\label{15}
\Sigma (s , \bar z\cdot z) =
\frac{2s - \kappa}{\pi}(1 -\kappa \bar z\cdot z)^{2s\kappa -2}.
\end{equation}
%where $r = \vert z \vert^2$.
One can write the function
$\psi(z)$ as the product of the state
$|\psi\rangle $ with some ket $\vert  \bar z \rangle$ labeled by the complex conjugates of the
variables $z$. This is
\begin{equation}\label{16}
\psi(z)= {\cal N} \langle  \bar z |\psi \rangle
\end{equation}
where ${\cal N}$ is a normalization constant to be adjusted later. Taking $|\psi\rangle = | s, n\rangle$, we
have
\begin{equation}\label{17}
\langle  \bar z\vert s, n\rangle = {\cal N}^{-1} C_{s,n}z^{n} .
\end{equation}
This leads to end up with the states
\begin{equation}\label{18}
| z \rangle =  {\cal N}^{-1}
\sum_{n}C_{s,n}
z^{n}|s, n\rangle
\end{equation}
which converges for
 $\kappa = +1$  when $\bar z\cdot z < 1$. Otherwise,
the Bargmann space coincides with the unit disc
 ${\cal D} =  \{ z \in {\mathbb C};\ \ \bar z \cdot z < 1 \}$.
The normalization constant is calculated as
\begin{equation}\label{19}
{\cal N} = (1 - \kappa \bar z\cdot z)^{-\kappa s}
\end{equation}
The states (\ref{18}) are continuous in the labeling and constitute
an overcomplete set, such that
\begin{equation}\label{20}
\int d\mu ( z , \bar z)\vert z \rangle \langle z \vert = \sum_n \vert s , n \rangle \langle s , n \vert
\end{equation}
 with respect to the measure
\begin{equation}\label{21}
d\mu ( z , \bar z) = d^2z {\cal N}^2\Sigma (s, \bar z\cdot z) =
d^2z \frac{2s - \kappa}{\pi}(1 -\kappa \bar z\cdot z)^{-2}.
\end{equation}
Therefore, they define an overcomplete set coherent states. The
Bargmann realization derived here turns out to be in one to one
correspondence with the lowest landau levels wavefunctions  for a
particle evolving in
 two-dimensional manifold in presence of a high strength magnetic field.
 This will be clarified in the next section by discussing
 the quantum mechanics on the manifold that we define by using
 the algebraic structures of the Weyl--Heisenberg algebra introduced above. In fact, we derive
 the abelian gauge field encoded in the geometry of the manifold
 and determine the wavefunctions in terms of Wigner functions, in
 particular that corresponding to the lowest
 energy level.

%%%%%%%%%%%%%%%%%%%%%%%%%%%%%%%%%%%%%%%%%%%%%%%%%%%%%%%%%%%%%%%
\section{Quantum mechanics on coset space ${\bf S}^2_{\kappa}$}
%%%%%%%%%%%%%%%%%%%%%%%%%%%%%%%%%%%%%%%%%%%%%%%%%%%%%%%%%%%%%%%

Using the generalized Weyl--Heisenberg algebra $W_{\kappa}$, we
define two-dimensional coset space ${\bf S}^2_{\kappa}$. We discuss
the quantum mechanics of a particle evolving in this space. In
quantizing this manifold, we write down the corresponding
wavefunctions and determine the associated spectrum.

%%%%%%%%%%%%%%%%%%%%%%%%%%%%%%%%%%%%%%%%%%%%%%%%%%%%%%
\subsection{Abelian connection and Wigner functions}
%%%%%%%%%%%%%%%%%%%%%%%%%%%%%%%%%%%%%%%%%%%%%%%%%%%%%

It is well established that for any Lie algebra ${\cal G}$, one can construct a geometrical manifold endowed with a symplectic structure.
This gives a phase space where the classical trajectories are defined. This manifold is isomorphic the so-called coset space $G/H$ where
$G$ is the covering group of the Lie algebra ${\cal G}$ and $H$ the maximal stability subgroup of $G$ with respect to a fixed state in the
representation space of the Lie algebra ${\cal G}$. More precisely, the manifold is obtained by an exponential mapping.
In two-dimensional curved space (with constant curvature), the suitable symmetry groups are $SU(1,1)$ and $SU(2)$. The algebra $W_{\kappa}$
provides us with an useful way to deal with these symmetries in a compact manner by defining  one parameter family of Lie
group $SU_{\kappa}(1,1)$ whose Lie algebra is generated by the infinitesimal generators $x_+$, $x_-$ and $2x_3 = {\mathbb I} + 2\kappa x_0$. Indeed, one can check from the commutation relations and the Hilbertian representations given in the previous section that
for $\kappa = +1$ , $W_{\kappa}$ reduces to the $su(1,1)$ algebra. %for $\kappa = 0$,  when
 However, in the case where
 $\kappa = -1$, we get the $su(2)$ Lie algebra. The maximal stability group is $U(1)$ and generated by the operator $x_3$. We shall focus on the algebra $W_{\kappa \neq 0}$. The limiting case $\kappa = 0$ can simply be recovered by a simple contraction procedure.

From the above considerations, one can introduce the coset space
%\begin{equation}\label{22}
${\bf S}^2_{\kappa} = SU_{\kappa}(1,1)/U(1)$
%\end{equation}
by carrying out an unitary exponential mapping
\begin{equation}\label{23}
\eta x_+ - \bar \eta x_- \longrightarrow  \exp (\eta x_+ - \bar \eta x_-)
\end{equation}
where $\eta$ is a complex variable.
The coset space $SU_{\kappa}(1,1)/U(1)$ is
generated by the  elements $g$,
which are  $2\times
2$ matrices of the fundamental representation of the group
$SU_{\kappa}(1,1)$. They satisfy the relation
\begin{equation}\label{24}
\det g = 1,\qquad \delta g^{\dag}
\delta = g^{-1}
\end{equation}
 with $\delta= {\rm diag}(1,-\kappa)$. An adequate parametrization of $g$ can be
 written as
\begin{equation}\label{25}
g= \pmatrix{\bar u_2&u_1\cr \kappa\bar u_1 & u_2\cr}
\end{equation}
where $u_1$ and $u_2$ are
the global coordinates of ${\bf S}^2_{\kappa}$, which they can be
mapped in terms of the local coordinates as
\begin{equation}\label{26}
u_1 = \frac{z}{\sqrt{1-\kappa \bar z \cdot z}}, \qquad u_2 =
\frac{1}{\sqrt{1-\kappa \bar z\cdot z}}.
\end{equation}

To write down the symplectic structure associated with the manifold ${\bf S}^2_{\kappa}$, we introduce
 the Maurer--Cartan one-form $g^{-1}dg $. A straightforward
 calculation gives
\begin{equation}\label{27}
g^{-1}dg = - i\ t_{+} \ e_{+} \ dz - i \ t_{-} \ e_{-}\ d\bar z - 2i\
\theta \ t_3
\end{equation}
where
 $t_{+}, t_{-}$ and $t_3$
are the $SU_{\kappa}(1,1)$ generators in the fundamental representation.
They  can be written in terms of
the   matrices $(e_{ij})_{kl} = \delta_{ik}\delta_{jl}$ of the
algebra  $gl(2)$ as
\begin{equation}\label{28}
t_{+} = -e_{12},\qquad
t_{-} = \kappa e_{21},\qquad  t_3 = \frac{1}{2}(e_{11} - e_{22}).
\end{equation}
The one-orthonormal forms $e_{+}$ and $e_{-}$, in (\ref{27}), reads as
\begin{equation}\label{29}
e_{+} = - \frac{i}{1-\kappa \bar z\cdot z}, \qquad e_{-}= \frac{i}{1-\kappa \bar
z \cdot z}.
\end{equation}
The $U(1)$ symplectic one-form, i.e. the $U(1)$ connection, is
\begin{equation}\label{30}
\theta = i \ {\rm Tr} \left(t_3g^{-1}dg \right) = \frac{i}{2}\
\frac{\bar z\cdot dz - z\cdot d\bar
z }{1 - \kappa \bar z\cdot z}
\end{equation}
which
 is the main quantity to  completely specify the geometry of   $SU_{\kappa}(1,1)/U(1)$
 and the corresponding symplectic structure.
With the above realization,  ${\bf S}^2_{\kappa}$
is equipped with the Kahler--Bergman metric
\begin{equation}\label{31}
 d\sigma^2 = \frac{1}{(1-\kappa\bar z\cdot z)^2} \ dz
d\bar z
\end{equation}
as well as a symplectic closed two-form
\begin{equation}\label{32}
\omega_0 = d\theta = (\omega_0)_{ \bar z\cdot z}\ dz \wedge d\bar z = \frac{i}{(1-\kappa\bar z\cdot z)^2}\ dz \wedge d\bar z.
\end{equation}

The manifold ${\bf S}_{\kappa}^2$ is symplectic and the associated Poisson bracket is given by
\begin{equation}\label{33}
\{f_1 , f_2 \} =  i(1-\kappa\bar z\cdot z)^2\left( \frac{\partial f_1}{\partial
z}\frac{\partial f_2}{\partial \bar z} - \frac{\partial f_1}{\partial
\bar z }\frac{\partial f_2}{\partial z}\right)
\end{equation}
where $f_1$ and $f_2$ are  two functions defined on  $SU_{\kappa}(1,1)$. They can be
expanded as
\begin{equation}\label{34}
f(g) = \sum_{n',n} f_{n',n}^s {\cal D}_{n',n}^s (g)
\end{equation}
and the Wigner ${\cal D}$-functions ${\cal D}_{n',n}^s (g)$
on $SU_{\kappa}(1,1)$ are defined by
\begin{equation}\label{35}
{\cal D}_{n',n}^s (g) = \langle s , n'| g |s , n\rangle.
\end{equation}
Recall that, $s$ is labeling the discrete  irreducible representations derived in the previous section.
These will allow us to define the wavefunctions of a system living in the coset space.

%%%%%%%%%%%%%%%%%%%%%%%%%%%%%%%%%%%%%%%%%%%%%%%%%%%%%%%%%
\subsection{Geometric quantization and Kahler vacuum }
%%%%%%%%%%%%%%%%%%%%%%%%%%%%%%%%%%%%%%%%%%%%%%%%%%%%%%%%

In the context of geometric quantization, the Kahler vacuum is
obtained from the wavefunctions (\ref{35}) by imposing the so-called
polarization condition, see equation (\ref{43}). This condition gives the
wavefunctions those are holomorphic (up to a normalization factor)
and coincide with LLL. % as it will be clarified in the next subsection.
For more details on the relation
between the polarisation and the projection on LLL, we refer for
instance to [23,24,26,27].

To show that the polarisation condition gives the coherent states
derived in subsection (2.3) and then defines the Kahler vacuum,
% To define the wavefunctions of a system living in the coset space ${\bf S}^2_{\kappa}$,
%we first look for the groundstate that corresponds to the Kahler
%vacuum. Indeed,
let us
 introduce the generators of the right $R_a$ and left $L_a$
translations of $g$, namely %. They are
\begin{equation}\label{36}
R_a g = g x_a,\qquad  L_a g = x_a g
\end{equation}
where  $ a $ runs for $= +, -, 3$. They act on the Wigner ${\cal D}$-functions as
\begin{equation}\label{37}
R_a {\cal D}_{n',n}^s (g) = {\cal D}_{n',n}^s (gx_a), \qquad L_a
{\cal D}_{n',n}^s (g) = {\cal D}_{n',n}^s (x_ag).
\end{equation}

To obtain the Kahler vacuum corresponding to  ${\bf S}_{\kappa}^2$,
we should reduce the
degrees of freedom on the manifold $SU_{\kappa}(1,1)$ to
the coset space $SU_{\kappa}(1,1)/U(1)$.  This reduction can be formulated
in terms of a suitable constraint on the Wigner ${\cal D}$-functions. Thus, we define the magnetic background by
\begin{equation}\label{38}
\omega = dA
\end{equation}
where the $U(1)$ gauge field potential is  proportional
to one-form (\ref{30}), i.e. $A=k\theta $ with $k$  the strength of the magnetic background  is a real number.
The suitable constraint on
the Wigner ${\cal D}$-functions can be
established by considering the $U(1)$ gauge transformation. This is
\begin{equation}\label{39}
g \rightarrow gh = g\exp(ix_3\varphi)
\end{equation}
where $\varphi$ is the $U(1)$ parameter. (\ref{39})
leads  to the transformation in the gauge field, such that
\begin{equation}\label{40}
A \rightarrow A + kd\varphi.
\end{equation}
It follows that the functions~(\ref{37}) transform as
\begin{equation}\label{41}
{\cal D}_{n',n}^s(gh) = \exp\left(\int\dot{A}dt\right){\cal D}_{n',n}^s (g)
= \exp\left(\frac{k}{2}\varphi\right){\cal D}_{n',n}^s (g).
\end{equation}
Therefore, the canonical momentum corresponding to the $U(1)$ direction
is $k/2$. Then, the admissible  states $\psi\equiv{\cal D}_{n',n}^s(g)$ must satisfy the
constraint
\begin{equation}\label{42}
R_3 \psi = \frac{k}{2} \psi.
\end{equation}
Thus, the physical states
constrained by~(\ref{42}) are  the Wigner ${\cal D}$-functions ${\cal D}_{n',n}^s
(g)$ where the quantum numbers are connected through the relation
$\frac{k}{2} = s + \kappa n$. To achieve the derivation of
the Kahler vacuum (groundstate), we use
the polarization condition:
\begin{equation}\label{43}
R_- {\cal D}_{n',n}^s (g) = 0
\end{equation}
which corresponds to $n=0$, i.e. $s=\frac{k}{2}$. This shows that the
index $s$, labeling
the $W_{\kappa}$ irreducible representations, is related to the
strength of the magnetic field. It is known in geometric quantization that the constraint (\ref{42}) and the polarization
condition (\ref{43}) define the Kahler vacuum. It corresponds to the functions
${\cal D}_{n',0}^s (g)$
\begin{equation}\label{44}
\psi_{\rm Kahler} \equiv {\cal D}_{n',0}^s (g)= \langle s,n'|g|s,0\rangle
\end{equation}
which are  holomorphic in the $z$-coordinate. More precisely,
in the fundamental representation,  we can define $g$
in terms of the generators $x_{\pm}$ by
\begin{equation}\label{45}
g = \exp(\eta x_+ - x_-\bar \eta)
\end{equation}
where $\eta$ are related to the local coordinates via
\begin{equation}\label{46}
z = \frac{\eta}{|\eta|}\tanh_{\kappa}|\eta| =
\frac{\eta}{\sqrt{\kappa}|\eta|}\frac{e^{\sqrt{\kappa}\eta} -
e^{-\sqrt{\kappa}\eta}}{e^{\sqrt{\kappa}\eta}+e^{-\sqrt{\kappa}\eta}}.
\end{equation}
Using (\ref{44}), we end up with the required functions
\begin{equation}\label{47}
\psi_{\rm Kahler} (\bar z, z) = (1-\kappa\bar
z\cdot z)^{\kappa\frac{k}{2}}C_{\frac{k}{2},n'}z^{n'}
\end{equation}
where $n' = 0, 1, \cdots , 2s$ for $\kappa = -1$ and $n' \in
{\mathbb N}$ for $\kappa = 1$. The Kahler vacuum is exactly the
Bargmann space constructed in the previous section. It is finite
(respectively infinite) dimensional for $\kappa = -1$ (respectively
$\kappa = 1$). Note that, the Kahler vacuum coincides with the
lowest landau levels of a particle evolving in the manifold ${\bf
S}^2_\kappa$, which will be clarified in the next subsection.

%%%%%%%%%%%%%%%%%%%%%%%%%%%%%%%%%%%%%%%%%%%%%%%%%%%%%%%%%%%%%%%%%%%%%%%%%%%%%%%%%%%%%%%%%%%%%%%%%%%%%%
\subsection{Energy spectrum solutions}
%%%%%%%%%%%%%%%%%%%%%%%%%%%%%%%%%%%%%%%%%%%%%%%%%%%%%%%%%%%%%%%%%%%%%%%%%%%%%%%%%%%%%%%%%%%%%%%%%%%%%

Since the manifold ${\bf S}^2_\kappa$ is constructed  from
the algebra $W_{\kappa}$, it is natural to consider the operator
\begin{equation}\label{48}
H_{\kappa} = \frac{1}{2}(R_+ R_- + R_-R_+)
\end{equation}
which generalizes the harmonic oscillator Hamiltonian.
 To establish a relation between  the Casimir operator
and the Hamiltonian, we may write the eigenvalue
equation as
\begin{equation}\label{49}
 H_{\kappa}\psi = \frac{1}{2}\left(x_-x_+ + x_+x_- \right)\psi = E \psi.
\end{equation}
Since the wavefunctions $\psi$ are the Wigner ${\cal D}$-functions ${\cal
  D}_{n', n}^s (g)$, satisfying the constraint $\frac{k}{2} = s + \kappa n$, the  energies are given by
\begin{equation}\label{50}
E^{\kappa} =  \frac{k^2}{4} - C_2
\end{equation}
where the second order Casimir operator is
\begin{equation}\label{51}
C_2 = x_3^2 - \frac{\kappa}{2}\left(x_-x_+ + x_+x_- \right).
\end{equation}
It is not difficult  to check that
the eigenvalues of $C_2$  are of the form $s(s-\kappa)$. Consequently, one can see that those
corresponding to (\ref{48}) are given by
\begin{equation}\label{52}
E := E^{\kappa}_n = {k\over 2}\left(2n+1\right)- \kappa n\left(n+1 \right).
\end{equation}
At this stage, we have some remarks in order. Indeed,
for $\kappa = \pm 1$, we recover the eigenvalues for one-particle living in the
disc and sphere geometries, respectively.
 %The Landau spectrum for the sphere is obtained from
%(52) when $\kappa = -1$.
Finally  for $\kappa = 0$, we have Landau spectrum on
two-dimensional Euclidean space. It is interesting to note also that,
the  Euclidian case can be obtained from (\ref{52}) for  $k$ large.
On the other hand, we notice here that the lowest level energy
corresponds to $n =0$. It is $2s+1$ fold generated for $\kappa = -1$ and infinitely
degenerated for $\kappa =1$. This is exactly the Kahler vacuum discussed above.

It is clear from (\ref{52}) that, for a large $k$, the spectrum rewrites as
\begin{equation}\label{53}
E_n^{\kappa} \sim \frac {k}{2}(2n+1)
\end{equation}
which is $\kappa$-independent and the gap between two successive levels, proportional to $k$,
is also large. In this situation,
the particles are constrained  to be accommodated in the lowest energy level. Since it is degenerated,
one may fill it with a large number of particles $N$ such that the density operator
is
%reads as
\begin{equation}\label{54}
\rho_0 = \sum_{n=0}^N \vert s , n \rangle \langle s, n \vert.
\end{equation}

Note that the quadratic Hamiltonian $H_{\kappa}$ involves only
the right translation. An admissible form for $H_{\kappa}$ should
also be expressed in terms of left translations. Since the right and
left generators commute, the operators that lift the degeneracies of
the energy levels  are functions of the left translations $L_a$.
Since we are interested in studying the excited states, we introduce
an excitation potential $V$ that induces a lifting of the LLL
degeneracy [23,26]. In this respect, it is natural to assume that
$V$ as function of the left translations $L_3$, $L_+$ and $L_-$. A
simple choice for $V$, in term of $L_3$, is given by
\begin{equation}\label{55}
V = \kappa \left(L_3 - \frac{k}{2}\right)
\end{equation}
Then, for $k$ large, the Hamiltonian
governing the dynamics of the system is now
 \begin{equation}\label{56}
H = E_0^{\kappa} + V.
\end{equation}
The corresponding eigenvalues are
\begin{equation}\label{57}
 H \psi_{{\rm Kahler}}\equiv  H {\cal D}_{n',0}^{\frac{k}{2}} (g)=
\left(\frac{k}{2} +  n'\right){\cal D}_{n',0}^{\frac{k}{2}} (g).
\end{equation}
This shows that we have a lifting of the degeneracy.

We close this section by noting that the lowest landau
wavefunctions constitute an overcomplete set.  One of the usefulness
of this property, see section (2), is that it provides us with a
simply way to establish a correspondence between operators and
classical functions on the phase space of the present systems. In
the next section, we develop a similar strategy to that  adopted in
[23,24,26,27]  to perform semiclassical calculations, which are
valid for high magnetic field.

%%%%%%%%%%%%%%%%%%%%%%%%%%%%%%%%%%%%%%%%%%%%%%%%%%%%%%%%%
\section{Semiclassical analysis and vacuum excitation}
%%%%%%%%%%%%%%%%%%%%%%%%%%%%%%%%%%%%%%%%%%%%%%%%%%%%%%%%%%
The dynamical description of a large collection of particles confined in the lowest Landau levels can be achieved semi classically.
Some tools are needed  in this sense. These concern the star product, the density function and the excitation potential.

%%%%%%%%%%%%%%%%%%%%%%%%%%%%%%%%%%%%%%%%%%%%%%%%%%%%%%%%%
\subsection{Star product}
%%%%%%%%%%%%%%%%%%%%%%%%%%%%%%%%%%%%%%%%%%%%%%%%%%%%%%%%%%

An important ingredient to perform the semiclassical
analysis in the Bargmann space is the star product. In fact, as we will discuss next,
for $s$ large the mean value of the product of two operators leads
to the Moyal star product. To show this, to every operator $O$
acting on the Fock space ${\cal F}$, we associate the function
\begin{equation}\label{58}
{\cal O}(\bar z, z) = \langle z | O | z \rangle.
\end{equation}
An associative star product of two functions ${\cal O}_1(\bar z, z)$
and ${\cal O}_2(\bar z, z)$ is defined by
\begin{equation}\label{59}
{\cal O}_1(\bar z, z)\star {\cal O}_2(\bar z, z) = \langle z | O_1O_2 | z
\rangle = \int d\mu(\bar{z'}, z') \langle z | O_1 | z' \rangle\langle z'| O_2 | z\rangle
\end{equation}
where the measure $d\mu(\bar z, z) = d^2z{\cal N}^2
\Sigma$, see (\ref{21}). To calculate this ,
let us exploit the analytical properties of coherent states
defined above.
 Indeed, using  (\ref{18}), one can see
that the function defined by
\begin{equation}\label{60}
{\cal {O}}(\bar z', z) = \frac{\langle z' | O | z \rangle}{\langle z'  | z \rangle}
\end{equation}
satisfies the holomorphic and anti-holomorphic conditions
\begin{equation}\label{61}
\frac{\partial}{\partial \bar{z}}{\cal {O}}(\bar z', z) = 0, \qquad
\frac{\partial}{\partial z'}{\cal {O}}(\bar z', z) = 0
\end{equation}
when $z\neq z'$. Consequently, the action of the translation operator
 on the function ${\cal {O}}(\bar z', z)$ gives %the result
\begin{equation}\label{62}
\exp\left(z'\frac{\partial}{\partial z}\right){\cal {O}}(\bar z', z) = {\cal {O}}(\bar z', z+z').
\end{equation}
This can be used to determine  ${\cal {O}}(\bar z, z')$ in terms of  ${\cal {O}}(\bar z, z)$. Indeed, we have
\begin{equation}\label{63}
\exp\left(-z \frac{\partial}{\partial z'}\right)\exp\left(z' \frac{\partial}{\partial z}\right)
{\cal {O}}(\bar z, z) =
 \exp\left((z'-z) \frac{\partial}{\partial z}\right){\cal {O}}(\bar z, z) = {\cal {O}}(\bar z, z').
\end{equation}
Similarly, one obtains
\begin{equation}\label{64}
\exp\left(-\bar z \frac{\partial}{\partial \bar z'}\right)
\exp\left(\bar z' \frac{\partial}{\partial \bar z}\right){\cal {O}}(\bar z, z) = {\cal {O}}(\bar z', z)
\end{equation}
Equivalently,  (\ref{63}) and (\ref{64}) can also be cast in the forms
\begin{equation}\label{65}
 \exp\left((z'-z) \frac{\partial}{\partial z}\right){\cal {O}}(\bar z, z) = {\cal {O}}(\bar z, z')
\end{equation}
as well as
\begin{equation}\label{66}
 \exp\left((\bar z'- \bar z) \frac{\partial}{\partial \bar z}\right){\cal {O}}(\bar z, z) = {\cal {O}}(\bar z', z).
\end{equation}
Combining (\ref{58}-\ref{60}) and (\ref{65}-\ref{66}), the star product rewrites as
\begin{equation}\label{67}
{\cal O}_1(\bar z, z)\star {\cal O}_2(\bar z, z) =
 \int d\mu(\bar{z'}, z') \exp\left((z'-z) \frac{\partial}{\partial z}\right){\cal {O}}_1(\bar z, z)
\vert\langle z | z' \rangle\vert^2\exp\left((\bar z'- \bar z) \frac{\partial}{\partial \bar z}\right){\cal {O}}_2(\bar z, z)
\end{equation}
where the overlapping of coherent states is given by
\begin{equation}\label{68}
\langle z' | z \rangle = \left[(1 -\kappa \bar z'\cdot z' )
(1 -\kappa \bar z z )(1 -\kappa \bar z'\cdot z )^{-2}\right]^{\kappa s}.
\end{equation}
Clearly, the modulus of the kernel (\ref{68}) possesses the properties $\vert\langle z | z' \rangle\vert = 1$
if and only if $z=z'$, $\vert\langle z | z' \rangle\vert<1$
 and $\vert\langle z | z' \rangle\vert \to 0$ for $k$ large $( k = 2s)$.
  These are helpful to get the star product between two functions
on the Bargmann space. In this respect, we introduce a function
$\sigma(z',z)$ of the coordinates of two points on the Bargmann space
\begin{equation}\label{69}
\sigma^2(z',z) = -{\rm ln} \vert \langle z | z' \rangle\vert^2 = 2\kappa  s\
{\rm ln} \frac{(1 -\kappa \bar z'\cdot z )(1 -\kappa \bar z\cdot z')}
{(1 -\kappa \bar z\cdot z )(1 -\kappa \bar z'\cdot z')}.
\end{equation}
It verifies the properties $\sigma(z',z) = \sigma(z,z')$ and $\sigma(z',z) = 0$ if and only if $z'=z$ . This
function defines the distance between two points $z$ and $z'$ on the Bargmann space. Indeed, one can verify that the line element
$d\sigma^2$, defined as the quadratic part of the decomposition of $\sigma^2(z, z+dz)$, gives (\ref{31}) up to a multiplicative factor.
% is given by
%\begin{equation}
%ds^2 =  g_{i\bar j} dz_id\bar z_j
%\end{equation}
%where summation over repeated indices is understood and the
%components of the metric $g_{i\bar j}$ are defined
%\begin{equation}
%g_{i\bar j} = (k + \frac{s}{2} - \frac{1}{2}) \bigg[\frac{\delta_{ij}}{1 -s \bar z.z } + s\frac{\bar z_iz_j}{(1 -s \bar z.z )^2}\bigg].
%\end{equation}

We now come to the evaluation of the star product for $k $ large. Since $\sigma^2(z',z)$ tends
to infinity with $k \to \infty$ if $z\neq z'$ and equals zero
if $ z = z'$, one can conclude that, in that limit, the domain $ z \simeq z'$ gives only
a contribution to the integral (\ref{67}). Decomposing the integrand near
the point $ z \simeq z'$ and going to integration over $\eta = z'-z$ , one gets
\begin{equation}\label{70}
{\cal O}_1(\bar z, z)\star {\cal O}_2(\bar z, z) =  \int \frac{d\eta d\bar{\eta}}{\pi}
 \exp\left(\eta \frac{\partial}{\partial z}\right){\cal {O}}_1(\bar z, z)
\exp\left(- i \omega_{\eta \bar{\eta}} \eta \bar{\eta}\right)
\exp\left(\bar{\eta} \frac{\partial}{\partial \bar z}\right){\cal {O}}_2(\bar z, z).
\end{equation}
It follows that the star product between two functions on the Bargmann space writes as
\begin{equation}\label{71}
{\cal O}_1(\bar z, z)\star {\cal O}_2(\bar z, z) = {\cal O}_1(\bar z, z)
{\cal O}_2(\bar z, z) - \frac{1}{k}( 1 - \kappa \bar z\cdot z)^2
\frac{\partial{\cal O}_1}{\partial{z}}(\bar z, z)\frac{\partial{\cal O}_2}{\partial{\bar z}}(\bar z, z)+
O\left(\frac{1}{k^2}\right).
\end{equation}
Then, the symbol or function associated with the commutator of two
operators $O_1$ and $O_2$ is given by
\begin{eqnarray}\label{72}
\langle z |[ O_1 , O_2] | z\rangle &=& \{{\cal O}_1(\bar z, z), {\cal O}_2(\bar z,
z)\}_{\star} \nonumber\\
 &=& - \frac{1}{k}( 1 - \kappa \bar z\cdot z)^2\left(
\frac{\partial{\cal O}_1}{\partial{z}}(\bar z, z)\frac{\partial{\cal O}_2}{\partial{\bar z}}(\bar z, z)
- \frac{\partial{\cal O}_2}{\partial{z}}(\bar z, z)\frac{\partial{\cal O}_1}{\partial{\bar z}}(\bar z, z)\right)
\end{eqnarray}
where the quantity
\begin{equation}\label{73}
 \{{\cal O}_1(\bar z, z), {\cal O}_2(\bar z,
z)\}_{\star} = {\cal O}_1(\bar z, z)\star {\cal O}_2(\bar z, z) -
{\cal O}_2(\bar z, z)\star {\cal O}_1(\bar z, z)
\end{equation}
is the so-called the Moyal bracket.

%%%%%%%%%%%%%%%%%%%%%%%%%%%%%%%%%%%%%%%%%%%%%%%%%%%%%%%%%
\subsection{Density matrix and excitation potential}
%%%%%%%%%%%%%%%%%%%%%%%%%%%%%%%%%%%%%%%%%%%%%%%%%%%%%%%%%%

The symbol function corresponding to the density operator (\ref{54})  is given by
\begin{equation}\label{74}
{\cal \rho }_0 (\bar z , z)= \langle z \vert \rho_0 \vert z \rangle = ( 1 - \kappa \bar z\cdot z)^{\kappa k}
\sum_{n=0}^N C_{n,\frac{k}{2}}^2 (\bar z\cdot z)^n.
\end{equation}
It is a simple matter of approximations to see that for $k$ and $N$ large, the classical density (\ref{74})
behaves like a circular droplet in the Bargmann space.
Indeed, one can see that (\ref{10}), for $k$ large, reduces to the form
\begin{equation}\label{75}
C_{n,\frac{k}{2}} \sim \frac{k^{\frac{n}{2}}}{\sqrt{n!}}.
\end{equation}
This leads to the density
\begin{equation}\label{76}
{\cal \rho }_0 (\bar z , z) \sim \exp(-k\bar z\cdot z)\sum_{n=0}^N \frac{(k \bar z\cdot z)^n}{n!}.
\end{equation}
It
 can be approximated by the step function, such that
\begin{equation}\label{77}
{\cal \rho }_0 (\bar z , z) = \Theta (N - k \bar z\cdot z)
\end{equation}
which is corresponding to a circular configuration in the Bargmann space with radius is proportional to $\sqrt N$.
The particles are confined in the interior of the disc $\{ z \in {\mathbb C},\ \  k\bar z\cdot z \leq N\}$.

The symbol associated
to the potential (\ref{55}) is given by
\begin{equation}\label{78}
 {\cal V}(\bar z, z ) = \langle z |V| z \rangle = k  \frac{\bar z\cdot z}{1-\kappa \bar z\cdot z}
\end{equation}
which behaves like  the harmonic oscillator potential for a strong
magnetic field. The star product (\ref{71}), Moyal bracket
(\ref{72}), density function (\ref{77}) and the excitation potential
(\ref{78}) are the main quantities needed in the derivation of the
action describing the electromagnetic excitation of two-dimensional
Hall droplets that we derive in the last section.

It is clear that, for $k$ large, a large collection of $N$
particles  behaves like Hall droplets of radius $N/k$. The Bargman
space (generated by the LLL wavefunctions) is the phase space of the
system. It is equipped with symplectic two-form $\omega$ (36) . The
dynamics of the system is described by the potential ${\cal V}$ with
%the symplectic two-form 
$\omega$. To describe the electromagnetic
excitations of Hall droplets,  we now consider the addition of a $U
(1)$ magnetic field described by a gauge potential $A$ such that the
total potential becomes $a + A$. The new dynamics is then given by
the symplectic form $\omega + F$. This induces noncommutative
structures in the Bargmann space, which will be discused
in the next section.

%%%%%%%%%%%%%%%%%%%%%%%%%%%%%%%%%%%%%%%%%%%%%%%%%%%%%%%
\section{Noncommutative dynamics in Bargmann space }
%%%%%%%%%%%%%%%%%%%%%%%%%%%%%%%%%%%%%%%%%%%%%%%%%%%%%%

At this stage, we show how the elecromagnetic excitation can be introduced from a purely symplectic scheme
and obtain the dynamics of the system by using (\ref{77}-\ref{78}). Recall that
it is well-known, in symplectic mechanics that, the coupling of a charged particle with an electromagnetic field can be
described in a Hamiltonian formalism without a choice of a gauge potential. This can be achieved through a modification
of the symplectic form. With this, the Poisson brackets become deformed leading to noncommutative structure like for instance
the noncommutative positions in the Landau problem. Such approach dealing with modified structure has been previously considered in connection
with quantum Hall effect [9,10]. It is also important to stress that due to Moser's lemma [12], it was
developed in [30,31], see also [32], the noncommutative gauge theory in curved spaces which  is essentially related to the
procedure of symplectic deformations.  In this section, we follows similar arguments as in [30,31] to study the symplectic deformation of Bargmann  space.

%%%%%%%%%%%%%%%%%%%%%%%%%%%%%%%%%%%%%%%%%%%
\subsection{{Deformed symplectic structure}}
%%%%%%%%%%%%%%%%%%%%%%%%%%%%%%%%%%%%%%%%%%

First, it is more convenient for our purpose to rewrite
 the symplectic two-form (\ref{38})
\begin{eqnarray}\label{79}
\omega = \frac{1}{2}\omega_{ij}(\xi) d\xi^i\wedge d\xi^j.
\end{eqnarray}
in terms of the real coordinates $\xi^1$ and $\xi^2$,  with $ z = (\xi^1 +   i \xi^2)/\sqrt{2}$.
To introduce the effect of a weak external electromagnetic interaction, let
 us consider the modified symplectic two-form. This is
\begin{eqnarray}\label{80}
\omega + F  = \frac{1}{2}\omega_{ij}(\xi) d\xi^i\wedge d\xi^j + \frac{1}{2}F_{ij}(\xi) d\xi^i\wedge d\xi^j,
\qquad i, j = 1, 2
\end{eqnarray}
where the closed electromagnetic tensor field is defined by
\begin{eqnarray}\label{81}
 F  = da, \qquad a = a_1(\xi^1 , \xi^2)d\xi^1 + a_2(\xi^1 , \xi^2)d\xi^2.
\end{eqnarray}
According to this symplectic deformation, the vector fields
\begin{eqnarray}\label{82}
 Y_{\cal F} = Y^i_{\cal F} \frac{\partial }{\partial \xi^i}
\end{eqnarray}
associated with a given function ${\cal F}(\xi^1 , \xi^2)$, are such that the interior contraction of $\omega + F$ with $ Y_{\cal F}$ gives
\begin{eqnarray}\label{83}
 \iota_ {Y_{\cal F}}(\omega + F ) = d{\cal F}.
\end{eqnarray}
After straightforward calculation, we obtain
\begin{eqnarray}\label{84}
 Y^i_{\cal F} = [(\omega + F)^{-1}]^{ji}\frac{\partial {\cal F}}{\partial \xi^j}
\end{eqnarray}
where  two-form $\omega + F$ is supposed invertible. The Poisson brackets corresponding
to this new electromagnetic background is now given by
\begin{eqnarray}\label{85}
 \{{\cal F} , {\cal G}\} = \iota _{Y_{\cal F}} \iota _{Y_{\cal G}}(\omega + F ) =
 [(\omega + F)^{-1}]^{ij}\frac{\partial {\cal F}}{\partial \xi^i}\frac{\partial {\cal G}}{\partial \xi^j}.
\end{eqnarray}
In particular, we obtain
\begin{eqnarray}\label{86}
 \{ \xi^1 , \xi^2\} = [(\omega + F)^{-1}]^{12}
 \end{eqnarray}
 which, for a weak electromagnetic field $F$, can also be written as
\begin{eqnarray}\label{87}
 \{ \xi^1 , \xi^2\} = (\omega^{-1})^{12} - (\omega^{-1})^{12}F_{21}(\omega^{-1})^{12} +\frac{1}{2}(\omega^{-1})^{12}F_{21}
 (\omega^{-1})^{12}F_{21}(\omega^{-1})^{12} + \cdots
 \end{eqnarray}
where $(\omega^{-1})^{ij}$ stand for matrix elements of the inverse of $\omega$. The
last two terms in (\ref{87}) encode the effect of the external magnetic excitation and
indicate the deformation of the symplectic structure of the Bargmann space. The equations of motion,
governing the dynamics of the system, read now
\begin{eqnarray}\label{88}
 (\omega + F)^{ij}\frac{d\xi^j}{dt}= \frac{\partial {\cal H}}{\partial \xi^i}.
 \end{eqnarray}
where ${\cal H} = \frac{k}{2} + {\cal V}$ and ${\cal V}$ is given in (\ref{78}).

%%%%%%%%%%%%%%%%%%%%%%%%%%%%%%%%%%%%%%%%%%%%%%%%%%%%%%%%%%%%%%%%%%%%%%%%%%%%%%%%%%%%%
\subsection{Symplectic dressing and Moser's lemma}
%%%%%%%%%%%%%%%%%%%%%%%%%%%%%%%%%%%%%%%%%%%%%%%%%%%%%%%%%%%%%%%%%%%%%%%%%%%%%%%%%%%%%

Moser's lemma [12], see also [30],  is essentially a refined
version of Darboux theorem according to which there always exists a
coordinate transformation to eliminate the electromagnetic force. In fact,
it provides a nice way to eliminate the fluctuations
induced by the electromagnetic field
 and is
realized by
 performing the dressing transformation  $\omega + F \longrightarrow \omega$. Thus, we have to  find
a diffeomorphism on the phase space $f$ relating $\omega$ and $\omega + F$ such as
\begin{eqnarray}\label{89}
f^{\star}(\omega + F) = \omega
\end{eqnarray}
{where $f^{\star}$ is the pullback that maps  $\omega + F$ into $\omega$, more details can be found in [30]}.  In this
respect, the Moser's lemma [12] constitutes the appropriate tool and
provides us with an elegant way to relate $\omega$ and $\omega + F$.
To explicit this, we start by considering a family of symplectic
forms
\begin{eqnarray}\label{90}
\omega(t) = \omega + t F
\end{eqnarray}
interpolating $\omega$ and $\omega + F$ for $t=0$ and $t=1$, respectively, with $ t\in [0,1]$. We also construct a family
of diffeomorphism
\begin{eqnarray}\label{91}
f^{\star}(t)\omega(t) = \omega
\end{eqnarray}
satisfying $f^{\star}(t=0)= id$ and the diffeomorphism $f^{\star}(t=1)$ will then be the required solution of our problem. In
order to find $f^{\star}(t)$, we introduce a $t$-dependent vector field $X(t)$ generating
 $f(t)$ as its flow. Differentiating the last equation, such that
 $X(t)$ has to satisfy the identity
\begin{eqnarray}\label{92}
0 = \frac{d}{dt} \left[f^{\star}(t)\omega(t)\right] = f^{\star}(t)\left[L_{X(t)}\omega(t) + \frac{d}{dt}\omega(t)\right]
\end{eqnarray}
where $L_{X(t)}$ denotes the Lie derivative of  $X(t)$. Using the Cartan
identity $L_{X} = \iota _X\circ d + d\circ \iota _X$
and the fact that $d\omega (t) = 0$, we have
\begin{eqnarray}\label{93}
 f^{\star}(t)\left[d\left\{(\iota _{X(t)}\omega(t)\right\} + F\right]=0.
\end{eqnarray}
Since $F = da$, it follows that  $X(t)$ has to satisfy the linear equation
\begin{eqnarray}\label{94}
\iota _{X(t)}\omega(t) + a = 0
\end{eqnarray}
which solves (\ref{93}). Is is easy to see that the components of  $X(t)$ are given by
\begin{eqnarray}\label{95}
X^i(t) = - a_j (\omega^{-1})^{ji} (t).
\end{eqnarray}
As we are dealing with weak electromagnetic perturbation, i.e. $F \ll \omega $,
 the matrix element in (\ref{95}) can be obtained from
\begin{eqnarray}\label{96}
 (\omega^{-1})(t) = \omega^{-1} - t(\omega^{-1}) F (\omega^{-1}) + t^2 (\omega^{-1}) F (\omega^{-1}) F (\omega^{-1}) + \cdots.
\end{eqnarray}

The $t$-evolution of $\omega(t)$ is governed by the first order differential equation
\begin{eqnarray}\label{97}
 \left[\partial _t + X(t)\right]\omega(t) = 0.
\end{eqnarray}
Thus,  two-forms $\omega(t+1)$ and $\omega(t)$ are related by
\begin{eqnarray}\label{98}
 [\exp(\partial _t + X(t))\exp(-\partial _t )]\omega(t+1) = \omega(t).
\end{eqnarray}
Finally, from the last equation, it is easy to see
\begin{eqnarray}\label{99}
 [\exp(\partial _t + X(t))\exp(-\partial _t )](t=0)(\omega + F) = f^{\star} (\omega + F)  = \omega
\end{eqnarray}
where the diffeomorphism ensuring the dressing transformation is then given by
\begin{eqnarray}\label{100}
f^{\star} = id + X(0) + \frac{1}{2} (\partial_t X)(0) + \frac{1}{2} X^2(0) +
O\left[(\omega^{-1})^3\right].
\end{eqnarray}
More explicitly, using (\ref{95}-\ref{96}), one obtains the contribution of the second term in (\ref{100}) as
\begin{eqnarray}\label{101}
X(0) = X^i(0)\partial_i = (\omega^{-1})^{ij}a_j\partial_i
\end{eqnarray}
which is $\omega^{-1}$-first order. The contribution of the third term in (\ref{100}) is
\begin{eqnarray}\label{102}
\frac{1}{2} (\partial_t X)(0) = -\frac{1}{2}\left[(\omega^{-1}) F (\omega^{-1})\right]^{ij}a_j\partial_i.
\end{eqnarray}
Similarly, the last term in (\ref{100}) is evaluated to get
\begin{eqnarray}\label{103}
\frac{1}{2} X^2(0) =
\frac{1}{2}\left[(\omega^{-1})^{ij}a_j\partial_i\right]
\left[(\omega^{-1})^{i'j'}a_{j'}\partial_{i'}\right].
\end{eqnarray}
Reporting the contributions (\ref{101}-\ref{103}) in (\ref{100}), one can
shows  that the diffeomorphism $f$ transforms the variables $\xi^l (l = 1, 2)$ as
\begin{eqnarray}\label{104}
f(\xi^l) = \xi^l + \xi^l_1 + \xi^l_2
\end{eqnarray}
where $\xi^l_1$ is given by
\begin{eqnarray}\label{105}
\xi^l_1 = (\omega^{-1})^{lj}a_j
\end{eqnarray}
and the last one reads as
\begin{eqnarray}\label{106}
\xi^l_2 = - \frac{1}{2}(\omega^{-1})^{lk} F_{kl'}(\omega^{-1})^{l'j}a_{j} +
\frac{1}{2}(\omega^{-1})^{ij}a_j\left[(\partial_i(\omega^{-1})^{lj'}\right]a_{j'}
+
\frac{1}{2}(\omega^{-1})^{ij}a_j(\omega^{-1})^{lj'}(\partial_ia_{j'}).
\end{eqnarray}
Using the relations
\begin{eqnarray}
&& \partial_{j'}a_{i'} = (\partial_{j'}\omega_{i'k'})\xi^{k'}_1 + \omega_{i'i}(\partial_{j'}\xi^{i}_1)\label{107}\\
&& \partial_i(\omega^{-1})^{lj'} = -(\omega^{-1})^{lj"}(\partial_i\omega_{j"k})(\omega^{-1})^{kj'}\label{107}
\end{eqnarray}
and the antisymmetry property of the symplectic form, one can verify
\begin{eqnarray}\label{109}
\xi^l_2 &=& -(\omega^{-1})^{lk}F_{kl'}(\omega^{-1})^{l'm}a_m %\nonumber\\
+ \frac{1}{2} (\omega^{-1})^{lk} (\omega^{-1})^{mj}a_j (\omega^{-1})^{nj'}a_{j'}\partial_m\omega_{nk}
\nonumber\\
&& + \frac{1}{2} (\omega^{-1})^{lk} (\omega^{-1})^{ms}a_s \omega_{mn} \partial_k\left[(\omega^{-1})^{ns'}a_{s'}\right].
\end{eqnarray}
 It is clear that the Moser's lemma is
very interesting in the symplectic dressing procedure and provides us with a
way to eliminate any fluctuation of the electromagnetic field strength by a simple
coordinates redefinition.

%%%%%%%%%%%%%%%%%%%%%%%%%%%%%%%%%%
\subsection{{Seiberg--Witten map}}
%%%%%%%%%%%%%%%%%%%%%%%%%%%%%%%%%%

It is interesting to note that, the dressing transformation based on Moser's lemma is behind
the Seiberg--Witten map. Indeed, one can see from (\ref{104}) that this transformation can
be written as
\begin{eqnarray}\label{110}
f(\xi^l) = \xi^l +  {\hat a}^l
\end{eqnarray}
where
\begin{eqnarray}\label{111}
{\hat a}^l &=& (\omega^{-1})^{lk} \ \bigg[a_k
- F_{kl'}(\omega^{-1})^{l'm}a_m+ \frac{1}{2}  (\omega^{-1})^{mj}a_j (\omega^{-1})^{nj'}a_{j'}\partial_m\omega_{nk}
\nonumber\\
&&+ \frac{1}{2}  (\omega^{-1})^{ms}a_s \omega_{mn} \partial_k((\omega^{-1})^{ns'}a_{s'})\bigg].
\end{eqnarray}
The relation (\ref{110}) is similar to the so-called Susskind map  introduced in the context of
noncommutative Chern--Simons theory in relation with the fractional quantum Hall effect [14]. It encodes the
geometrical fluctuations induced by the external magnetic field $F$.
More importantly, (\ref{110}) coincides with the Seiberg--Witten map
in a curved manifold  for an abelian gauge theory [30].

Under the gauge transformation
\begin{eqnarray}\label{112}
a \longrightarrow a + d\lambda
\end{eqnarray}
 the components (109) transform as noncommutative abelian gauge field
 \begin{eqnarray}\label{113}
{\hat a}^l \longrightarrow {\hat a}^l + (\omega^{-1})^{lj}\partial_j {\hat \lambda} + \{ {\hat a}^l , {\hat \lambda}\}+ \cdots
\end{eqnarray}
where  the noncommutative gauge parameter ${\hat \lambda}$
\begin{eqnarray}\label{114}
{\hat \lambda} = \lambda + \frac{1}{2} (\omega^{-1})^{ij}a_j\partial_i\lambda + \cdots
\end{eqnarray}
is written in terms of the gauge parameter $\lambda$ and the $U(1)$ connection $a$.
It is  clear that
the dressing transformation, using the Moser's lemma, induces a noncommutative gauge field.
This establish a nice correspondence between the symplectic deformations and noncommutative gauge theories.

%%%%%%%%%%%%%%%%%%%%%%%%%%%%%%%%%%%%%%%%%%%%%%%%%%%%%%%%%%%%%%%%
\subsection{Hamiltonian and induced Chern--Simons term }
%%%%%%%%%%%%%%%%%%%%%%%%%%%%%%%%%%%%%%%%%%%%%%%%%%%%%%%%%%%%%%%%

As discussed above, we are interested in the  droplets (\ref{77}) on the manifold ${\bf S}^2_{\kappa}$ equipped with the
symplectic two-form $\omega$. The lowest energy levels are described by the quantization of
Bargmann space  with this symplectic form. In the
situation where the system evolves under the action of an external electromagnetic interaction $F = da$, the dynamics of the system becomes governed by $\omega + F$ together with the Hamiltonian
${\cal H} + a_0$, with ${\cal H}$ is the symbol of $H$ given by (\ref{56}) and
$a_0$ is the time component of the gauge potential $a$. In this case, a complete description of the dynamics is
encoded in the couple $(\omega + F, {\cal H} + a_0)$ and the equations of motion are given by (\ref{88}).
The Moser's lemma provides us with
a nice tool to incorporate the external interaction in the Hamiltonian leaving the symplectic form $\omega$ unchanged. Indeed,
the dressing transformation (\ref{104}) allows us to describe the dynamics of the system with the couple $(\omega , {\cal H}_a )$
 where we use the old symplectic but a new Hamiltonian. It is
\begin{eqnarray}\label{115}
{\cal H}_a (\xi^1, \xi^2)= ({\cal H}+ a_0)\left[f(\xi^1), f(\xi^2)\right].
\end{eqnarray}
This can be expanded, up to second order, as
\begin{eqnarray}\label{116}
{\cal H}_a (\xi^1, \xi^2)= {\cal H}+ a_0 + \xi_1^i\partial_i({\cal H}+ a_0) + \frac{1}{2}
\xi_1^i\xi_1^j\partial_i\partial_j ({\cal H}+ a_0) + \xi_2^i\partial_i({\cal H}+ a_0).
\end{eqnarray}
The dynamics of the system is now governed by
the equations of motion
\begin{eqnarray}\label{117}
 (\omega^{-1} )^{ij}\frac{d\xi^j}{dt}= \frac{\partial {\cal H}_a}{\partial \xi^i}.
 \end{eqnarray}
This shows that the dressing transformation, based on Moser's lemma, eliminates the fluctuations
in the symplectic structure and incorporate the electromagnetic interaction effect in the Hamiltonian. In other
words,
 this means that the dynamics, governed by  $\omega + F$ and  $ {\cal H}$, can
 be described by the old (non-deformed) symplectic two-form $\omega$ with a new Hamiltonian ${\cal H}_a$ expressed
in terms of the electromagnetic field $F$.

Injecting (\ref{105}) and (\ref{109}) in (\ref{116}) and using the expression of the excitation potential (\ref{78}),
one shows that the new Hamiltonian takes the form
\begin{eqnarray}\label{118}
{\cal H}_a &=& {\cal H} + a_0 + \epsilon^{ij}a_i\xi_j + \frac{1}{2}(\omega^{-1})^{ij}
\partial_i \left[(a_0 + \epsilon^{kl}a_l\xi_k)a_j\right]\nonumber \\
&&+ \frac{1}{2}(\omega^{-1})^{ij}\left[ a_i\partial_0a_j - 2 a_i\partial_ja_0 - \partial_i(a_0a_j)\right]
\end{eqnarray}
where we added $\frac{1}{2}(\omega^{-1})^{ij}a_i\partial_0a_j$ to ensure the gauge invariance under the change $a_0 \to
a_0 + \partial_0\lambda$.
Using the expressions of  $\omega$, one can rewrite the expression (\ref{118}) as
\begin{eqnarray}\label{119}
{\cal H}_a = {\cal H} + a_0 + \epsilon^{ij}a_i\xi_j - \frac{1}{2\sqrt{{\rm det}\omega}}\left\{
\epsilon^{ij}\partial_i \left[(a_0 + \epsilon^{kl}a_l\xi_k)a_j\right]
- \epsilon^{\alpha \beta \gamma}a_{\alpha}\partial_{\beta} a_{\gamma}\right\}
\end{eqnarray}
where $\alpha, \beta,\gamma = 0, 1, 2$.
This form is more suggestive because it  involves  a Chern--Simons term, i.e.
 last contribution in (\ref{119}).

%%%%%%%%%%%%%%%%%%%%%%%%%%%%%%%%%%%%%%%%%%%%%%%%%%%%%%%%%%%%%%%%%%%%%%%%%%%%%%%%%%%%%%%%%%%%%%%%%%%%%%
\section{
Electromagnetic excitations of quantum Hall droplet}
%%%%%%%%%%%%%%%%%%%%%%%%%%%%%%%%%%%%%%%%%%%%%%%%%%%%%%%%%%%%%%%%%%%%%%%%%%%%%%%%%%%%%%%%%%%%%%%%%%%%%

The analysis developed in the previous section can be applied to derive the electromagnetic
interaction of edge excitations of a two dimensional Hall system.

%%%%%%%%%%%%%%%%%%%%%%%%%%%%%%%%%%%%%%%%%%%%%%%%%%%%%%%%%%%%%%%%%%%%%%%%%%%%%%%%%%%%%%%%%%%%%%%%%%%%%%
\subsection{Effective WZW action}
%%%%%%%%%%%%%%%%%%%%%%%%%%%%%%%%%%%%%%%%%%%%%%%%%%%%%%%%%%%%%%%%%%%%%%%%%%%%%%%%%%%%%%%%%%%%%%%%%%%%%
Once we determined the spectrum of the lowest Landau levels where
the quantum Hall droplet is specified by the density matrix
$\rho_0$, one may ask about the excited states. The answer can be
given by describing the excitations in terms of an unitary time
evolution operator $U$. It contains  information concerning the
dynamics of the excitations around $\rho_0$. Therefore the excited
states will be characterized by a density operator given by
\begin{equation}\label{120}
{\rho} = U {\rho}_0 U^{\dag}.
\end{equation}
This is basically corresponding to a perturbation of the quantum system.

Thus, the
dynamical information, related to the degrees of freedom of the edge
states, is contained in the unitary operator $U$. The action,
describing these excitations, in the Hartree--Fock approximation,
can be written as [33]
\begin{equation}\label{121}
S = \int dt{\rm  Tr}\ \left( i\rho_0 U^{\dag} \partial_t U - \rho_0 U^{\dag} H_aU \right)
\end{equation}
where $H_a$ is  the operator associated to the Hamiltonian function given by~(\ref{119}). For a strong magnetic
field, i.e. large $k$, the different quantities occurring in the action can be
evaluated as classical functions. To do this, we adopt a
method similar to that used in [23]. This is mainly based on the strategy
elaborated by Sakita [33] in dealing with a bosonized theory for
fermions.

To determine the effective action,
we start by calculating the kinetic contribution, i.e. the first term
in r.h.s. of~(\ref{121}). This can be done by setting
\begin{equation}\label{122}
U= e^{+i\Phi}, \qquad
\Phi^{\dag} = \Phi.
\end{equation}
Using the definition of star product (\ref{59}), it is a matter of computation to see that
\begin{equation}\label{123}
 i \int dt {\rm Tr}\left(\rho_0 U^{\dag}\partial_tU \right) \simeq \frac
 {1}{2k} \int d\mu( \bar z, z )
\ \{\phi,\rho_0\} \ \partial_t\phi
\end{equation}
where we have dropped the terms in $\frac{1}{k^2}$ as well as the total time derivative. In (\ref{123}),
$\phi$ stands for the classical function associated to the operator $\Phi$.

The Poisson bracket can be calculated to get
\begin{equation}\label{124}
\{\phi , \rho_0\} = ({\cal L}\phi)  \frac{\partial\rho_0}{\partial
(\bar z\cdot z)}
\end{equation}
and  the first order differential operator ${\cal L}$ is given by
\begin{equation}\label{125}
 {\cal L} =  i \left(1 - \kappa \bar z\cdot z\right)^2 \left(z\cdot \frac{\partial}{\partial z}
- \bar z\cdot \frac{\partial}{\partial \bar z}\right).
\end{equation}
This is the angular momenta mapped in terms of the local coordinates
of ${\bf S}^2_{\kappa}$.
Recall that, for large $k$, the density~(\ref{77}) is a step
function. Its derivative is a $\delta$-function with a support on
the boundary $\partial {\cal D} ={\bf S}^1$ of the quantum Hall droplet ${\cal D}$. By setting $z = re^{i\alpha}$,
we show that~(\ref{125}) reduces to $ {\cal L} =
\partial_{\alpha}$ for large $k$. Therefore,
(\ref{123}) takes the form
\begin{equation}\label{126}
 i \int dt {\rm Tr}\left(\rho_0 U^{\dag}\partial_tU \right) \approx
 -\frac{1}{2} \int_{{\bf S^1}\times{\bf
R^+}} dt \ \left(\partial_{\alpha}\phi\right) \left(\partial_t\phi \right).
\end{equation}

Now we come to the semiclassical evaluation of the potential energy term. i.e. the second term in r.h.s. of~(\ref{121}).
By a straightforward calculation, we find
\begin{equation}\label{127}
{\rm  Tr}\left(\rho_0 U^{\dag}  H_a U\right) = {\rm  Tr}\left(\rho_0   H_a \right) + i
{\rm  Tr}\left(\left[\rho_0, H_a\right] \Phi\right)
+ \frac{1}{2}\  {\rm Tr}\left(\left[\rho_0, \Phi \right] \left[ H_a, \Phi
  \right] \right).
\end{equation}
The term in r.h.s of~(\ref{127}) is $\Phi$-independent. It is given by
\begin{equation}\label{128}
{\rm  Tr}\left(\rho_0  H_a \right)= \int d\mu(\bar z, z) \ \rho_0  \star {\cal H}_a.
\end{equation}
This term gives no contribution when $a=0$. The second term in (\ref{127}) can be written in term of the
Moyal bracket as
\begin{equation}\label{129}
 i {\rm  Tr}\left(\left[\rho_0, H_a\right] \Phi\right) \approx -\frac{1}{k}
\int d\mu(\bar z, z) ({\cal L}\phi)  \frac{\partial\rho_0}{\partial
r}{\cal H}_a.
\end{equation}
%in term of the Moyal bracket which, in the limit $n$ large, tends to
%the Poisson bracket (see (72)).
This term gives no contribution in the absence of electromagnetic interaction.
The last term in r.h.s of~(\ref{127}) can be evaluated in a similar
way to get~(\ref{129}).  One obtains
\begin{equation}\label{130}
\frac{1}{2}\  {\rm Tr}\left(\left[\rho_0, \Phi \right] \left[  H_a , \Phi
  \right] \right) = - \frac{1}{2k^2}\int d\mu(\bar z, z)\ \left({\cal L
}\phi\right)\ \frac{\partial\rho_0}{\partial r}\ \left({\cal L}\phi\right)
\ \frac{\partial {\cal V}}{\partial r}.
\end{equation}
Note that, we have eliminated a term containing the ground state
energy $E^{\kappa}_0$, because does not contribute to the edge dynamics. Also  we ignored in the last
equation the contributions coming from the electromagnetic interaction.

The addition of the contributions (\ref{123}) and (\ref{130}), which are $a$-independents, gives
\begin{equation}\label{131}
S_0 = \frac{1}{2k}\int dt\ d\mu(\bar z, z)\ \frac{\partial\rho_0}{\partial r}\bigg[\left({\cal L
}\phi\right) (\partial_t\phi) + \frac{1}{k}
\ \frac{\partial {\cal V}}{\partial r}\left({\cal L}\phi\right)^2\bigg].
\end{equation}
This is the Wess--Zumino--Witten action describing the edge
excitations of quantum Hall droplets in two dimensional space
[33-34].

%%%%%%%%%%%%%%%%%%%%%%%%%%%%%%%%%%%%%%%%%%%%%%%%%%%%%%%%%%%%%%%%%%%%%%%%%%%%%%%%%%%%%%%%%%%%%%%%%%%%%%
\subsection{Total action}
%%%%%%%%%%%%%%%%%%%%%%%%%%%%%%%%%%%%%%%%%%%%%%%%%%%%%%%%%%%%%%%%%%%%%%%%%%%%%%%%%%%%%%%%%%%%%%%%%%%%%

The total action is then given by
\begin{equation}\label{132}
S = S_0 + S_a
\end{equation}
where the $S_a$ part
\begin{equation}\label{133}
S_a = - \int dt\ d\mu(\bar z, z)\ \bigg[ \rho_0  \star {\cal H}_a - \frac{1}{k} ({\cal L}\phi)  \frac{\partial\rho_0}{\partial
r}{\cal H}_a\bigg]
\end{equation}
is the sum of
(\ref{128}) and (\ref{129}) containing the effect of the electromagnetic interaction. More precisely,
it is composed of the edge and bulk
contributions, such that
\begin{equation}\label{134}
S_a = S_a^{\rm bulk} + S_a^{\rm edge}
\end{equation}
where the bulk action
\begin{equation}\label{135}
S_a^{\rm bulk} = - \int dt\ d\mu(\bar z, z)\  \rho_0  \star \bigg[{\cal H}_a - \frac{1}{2}(\omega^{-1})^{ij}
\partial_i \bigg((a_0 + \epsilon^{kl}a_l\xi_k)a_j\bigg)\bigg]
\end{equation}
is containing a Chern--Simons action. The edge contribution, arising from the electromagnetic excitations,
reads as
\begin{equation}\label{136}
S_a^{\rm edge} = - \int dt\ d\mu(\bar z, z)\ \bigg[ \frac{1}{2}\rho_0  \star \bigg((\omega^{-1})^{ij}
\partial_i [(a_0 + \epsilon^{kl}a_l\xi_k)a_j] \bigg)- \frac{1}{k} ({\cal L}\phi)  \frac{\partial\rho_0}{\partial
r}{\cal H}_a\bigg].
\end{equation}
It is clear that, the second term in the above action is a boundary contribution. i.e. the derivative of the density
gives a delta function defined on the edge of the quantum Hall droplet. It is also easy to show that the first term
in (\ref{136}) is an edge contribution. Indeed, using the expression of the measure (\ref{21}) and the
inverse matrix elements of the two-form $\omega$,  one can verify
\begin{equation}\label{137}
\int dt\ d\mu(\bar z, z)\  \frac{1}{2}\rho_0  \star (\omega^{-1})^{ij}
\partial_i \left((a_0 + \epsilon^{kl}a_l\xi_k)a_j\right) \sim   \int dt\ d^2\xi \epsilon^{ij}\xi_ia_j(a_0 + \epsilon^{kl}a_l\xi_k)a_j)\frac{\partial\rho_0}{\partial
r}+ \cdots.
\end{equation}

It is remarkable that for both
cases $\kappa = 1$, $\kappa = 0$ and $\kappa = -1$ corresponding
respectively to hyperbolic, Euclidean and spherical geometry, we obtain the same expression for the
WZW action. This is mainly due to the fact
that we considered a large magnetic field so that the particles are constrained to be in the lowest landau levels. More
importantly, by reporting the Hamiltonian ${\cal H}_a$ given by (\ref{119}) in the action $S_a$ (\ref{133}), it is easily verified that
  the action $S$ agrees with the result derived in [27]. This corroborates our  claim according to which the
electromagnetic coupling of a quantum Hall droplet can be described by a deformation of the symplectic structure of Bargmann phase space
associated with the lowest landau levels. In such  description, the key tool is Moser's lemma which permits to incorporate the interaction effect in the Hamiltonian. The present application gives a simple way
to obtain the effective action describing the electromagnetic interaction of Kahler vacuum or quantum Hall droplets for others geometries.

Finally, notice that in the absence of electromagnetic field $ a = 0$ and for large $k$, from~(\ref{132})
we recover the effective WZW action describing the edge excitations of a dense collection
of fermions in two-dimensional space [33,34]. Otherwise for $a = 0$, the action $S$ reduces to $S_0$.

%\newpage

%%%%%%%%%%%%%%%%%%%%%%%%%%%%%%%%%%%%%%%
\section{Concluding remarks}
%%%%%%%%%%%%%%%%%%%%%%%%%%%%%%%%%%%%%%%%

We introduced a generalized Weyl--Heisenberg $W_{\kappa}$ (in fact a one parameter
family of algebras) that includes the harmonic oscillator, $su(2)$ and $su(1,1)$ algebras.
We constructed the corresponding Fock--Bargmann phase space, which provided us with an useful
tool to perform semiclassical analysis.  We also defined a Lie group
$SU_{\kappa}(1,1)$ involving the aforementioned symmetries and a homogeneous space,
 which also includes the three
two-dimensional surfaces (plane, sphere, disc) where the quantum system lives in.
In quantizing the coset space $SU_{\kappa}(1,1)/U(1)$, we showed that the Landau quantum
(or integer quantum Hall) systems in a plane $\kappa = 0$, sphere $\kappa = -1$ and disc $\kappa = 1$
 can be described in an unified algebraic scheme
using the algebra $W_{\kappa}$. This unified formulation provides also a
nice way to study the electromagnetic excitations of quantum Hall droplets. This is done from a
purely symplectic point of view by modifying the symplectic structure of underlying phase spaces.

Subsequently, we showed that through a dressing transformation, based on the Moser's lemma that
is a refined version of the celebrated Darboux theorem, one can find a diffeomorphism which
eliminate the fluctuation. In this way, the electromagnetic interaction becomes incorporated in
the Hamiltonian involving a Chern--Simons like term. Note also the deep relation between
Moser's lemma and Seiberg--Witten map. Finally, we gave the effective action
governing the electromagnetic excitations of the quantum Hall droplets.

The results of the present work can be extended in many directions. For
instance, we may study the higher dimensional phase spaces. Indeed, one can
generalize the obtained results so far to a quantum system living in four-dimensional
phase space. In this case, the modification
of the symplectic two-form, the position and momentum variables acquire nonvanishing
Poisson brackets inducing upon quantization non commutative positions and momenta operators. On the other
hand, in four-dimensional case, for some particular forms of the electromagnetic two-form, it is possible
to realize the dressing transformation by help of the so-called Hilbert-Schmidt orthonormalisation procedure . Thus,
it will be interesting to compare this method with that based on the Moser lemma.

In an other context, the phase space description
of internal degrees of freedom of particles obeying $A_r$ statistics [35,36] are described
by a $2r$-dimensional  Bargmann space [37]. The present work gives the main tools to deal with the electromagnetic
interaction of such systems in the semiclassical regime. We hope to report on these issues in a forthcoming work.

%%%%%%%%%%%%%%%%%%%%%%%%%%%%%%%%%%%%%%
\section*{ Acknowledgments}
%%%%%%%%%%%%%%%%%%%%%%%%%%%%%%%%%%%%%%

MD would like to thank the hospitality and kindness of
Max Planck Institute for Physics of Complex Systems (Dresden-Germany)
and Abdus Salam International
 Centre for Theoretical Physics (Trieste-Italy). AJ is grateful to
 Dr. Abdullah Aljaafari for his help and support.

\end{document}